\documentclass{article}

\usepackage{arxiv}

\usepackage[utf8]{inputenc} 
\usepackage[T1]{fontenc}    
\usepackage{hyperref}       
\usepackage{url}            
\usepackage{booktabs}       
\usepackage{amsfonts}       
\usepackage{nicefrac}       
\usepackage{microtype}      
\usepackage{lipsum}		
\usepackage{graphicx}
\usepackage[square , sort, numbers]{natbib}
\usepackage{doi}
\usepackage[nolist]{acronym}

\title{Testbed and Software Architecture for Enhancing Security in Industrial Private 5G Networks}

\date{July 28, 2025}	

\author{ \href{https://orcid.org/0000-0003-1716-5867}{\includegraphics[scale=0.06]{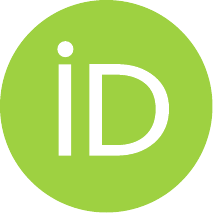}\hspace{1mm}Song Son Ha}\\
	Electrical Measurement Engineering\\
	Helmut-Schmidt-University\\
	Hamburg, Germany\\
	\texttt{song.ha@hsu-hh.de} \\
    \And
	\href{https://orcid.org/0009-0003-4437-010X}{\includegraphics[scale=0.06]{orcid.pdf}\hspace{1mm}Florian Foerster}\\
	Institute for Innovative Safety and Security\\
	Technical Unviersity of Applied Sciences Augsburg\\
	Augsburg, Germany\\
	\texttt{florian.foerster@tha.de} \\
	\And
	\href{https://orcid.org/0000-0002-6882-1214}{\includegraphics[scale=0.06]{orcid.pdf}\hspace{1mm}Thomas Robert Doebbert}\\
	Electrical Measurement Engineering\\
	Helmut-Schmidt-University\\
	Hamburg, Germany\\
	\texttt{thomas.doebbert@hsu-hh.de} \\
	\And
	{\hspace{1mm}Tim Kittel}\\
	ipoque GmbH\\
	A Rohde \& Schwarz company\\
	Leipzig, Germany\\
	\texttt{tim.kittel@rohde-schwarz.com} \\
	\And
	\href{https://orcid.org/0000-0003-2310-5895}{\includegraphics[scale=0.06]{orcid.pdf}\hspace{1mm}Dominik Merli}\\
	Institute for Innovative Safety and Security\\
	Technical Unviersity of Applied Sciences Augsburg\\
	Augsburg, Germany\\
	\texttt{florian.foerster@tha.de} \\
	\And
	{\hspace{1mm}Gerd Scholl}\\
	Electrical Measurement Engineering\\
	Helmut-Schmidt-University\\
	Hamburg, Germany\\
	\texttt{gerd.scholl@hsu-hh.de} \\
}

\renewcommand{\shorttitle}{Testbed and Software Architecture for Enhancing Security in Industrial Private 5G Networks}

\hypersetup{
pdftitle={Testbed and Software Architecture for Enhancing Security in Industrial Private 5G Networks},
pdfsubject={Security, Private 5G, DPI, Machine Learning},
pdfauthor={Song Son Ha},
pdfkeywords={Security, Private 5G, DPI, Machine Learning},
}

\begin{document}
\maketitle


\begin{abstract}
\footnote{This is the author's version of a paper that has been accepted for presentation at the 30th IEEE International Conference on Emerging Technologies and Factory Automation (ETFA 2025), to be held in Porto, Portugal, on September 9–12, 2025.}
In the era of Industry 4.0, the growing need for secure and efficient communication systems has driven the development of fifth-generation (5G) networks characterized by extremely low latency, massive device connectivity and high data transfer speeds. However, the deployment of 5G networks presents significant security challenges, requiring advanced and robust solutions to counter increasingly sophisticated cyber threats. This paper proposes a testbed and software architecture to strengthen the security of Private 5G Networks, particularly in industrial communication environments.
\end{abstract}

\acresetall


\section{Introduction}

The rapid development of industrial automation, driven by the Industry 4.0 revolution, has significantly increased the need for secure and reliable communication networks. The 5G technology, characterized by ultra-low latency and high reliability, also sets a focus on data security requirements, addressing the special needs of industrial environments \cite{Integration5GOPCUA, private5g}. While offering high flexibility, wide-area connectivity, and rapid deployment \cite{private5g}, these properties also increase the attack surface, posing risks to industrial operations, data integrity, and system availability \cite{Securitythread}. This risk becomes particularly critical in environments with functional safety applications, where any irregularity could threaten both operational availability and human, machine, and environmental safety \cite{doebberttestbed, HenryMeasurements}.
To secure conventional networks, firewalls and traditional Intrusion Detection Systems (IDS) are commonly employed. Nevertheless, their protection is often insufficient against advanced and evolving threats \cite{IDSFirewall}. In modern industrial environments such as Private 5G Networks, there is a critical need to develop more sophisticated and adaptive security solutions. In this context, several publications studied the integration of Deep Packet Inspection (DPI) with Machine Learning (ML) as a promising approach for monitoring and analyzing network traffic \cite{ResearchDPIML,transformerstein}. However, the effectiveness of this approach in securing industrial Private 5G Networks has still to be proven.	Moreover, the limited availability of real-world network traffic datasets from industrial environments poses a significant challenge for training and evaluating ML models \cite{ML_Based_IoT}. Therefore, it is essential to develop a testbed and software architecture that enables both the development of advanced DPI- and ML-based security solutions and the generation of high-quality, tailored industrial datasets.

This paper is structured as follows: Section~\ref{chap:RelatedWork} provides related work. Section~\ref{chap:Architecture} outlines the proposed architecture, while Section~\ref{chap:Testbed} describes its testbed implementation in detail. In Section~\ref{chap:Evaluation}, the current results of testbed experiments are presented. Finally, Section~\ref{chap:Conclusions} concludes the paper and outlines future work.

\section{Related Work} \label{chap:RelatedWork}

Wen et al. \cite{VET5G} introduce a virtual testbed (VET5G) for 5G security research using OpenAirInterface and Android emulators to simulate end-to-end networks, although their emulation-only approach limits applicability to industrial environments.
Baccar et al. \cite{TestbedBaccar} propose a testbed for real-time 5G packet generation and injection for security evaluation with a focus on deploying attacks to facilitate fuzzing-based vulnerability detection rather than implementing an intrusion detection approach.
Similarly, Almazyad et al. \cite{almazyad} present a 5G testbed that combines open-source software and hardware, focusing solely on simulating and analyzing cyberattacks such as DoS and database exploits.
Storm et al. \cite{Storm2024} design a test environment for evaluating existing industrial IDS solutions, but do not include a custom implementation, unlike our tailored approach.

Yang et al. \cite{ResearchDPIML} propose a DPI- and ML-based method for monitoring and identifying encrypted or unknown traffic. However, their work is limited to algorithm-level evaluation without deployment in realistic testbeds or relevance to industrial Private 5G Networks.
Stein et al. \cite{transformerstein} propose a transformer-based DPI algorithm for malicious traffic detection, while Bindra et al. \cite{Bindra2024} develop an IDS for OPC UA traffic focusing on DoS attacks. However, both studies rely on public datasets and lack a real-world test environment.	
Similarly, Jonghoon et al. \cite{Jonghoon} apply deep learning to 5G traffic intrusion detection using public datasets and evaluate their solution in a model factory, but the lack of a dataset tailored to their environment limits its specificity and suitability.	

While prior studies offer valuable contributions to 5G security through testbed designs, DPI- and ML-based techniques, they typically either remain confined to simulated environments, focus solely on attack deployment, or rely on public datasets without validation of realistic industrial 5G settings. To address these gaps, our work presents an testbed and software architecture that supports the development of advanced IDS based on DPI and ML, along with the generation of customized datasets tailored to industrial Private 5G Networks.

\begin{figure}[tb]
	\centering
	\includegraphics[width = 1\columnwidth]{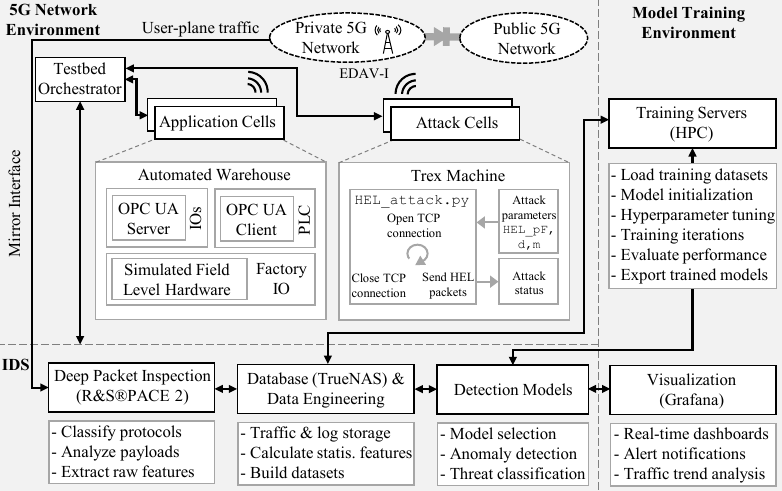}
	\caption{Overview of the proposed architecture with key functions and selected implementations} 
	\label{architecture} 
\end{figure}

\section{Proposed Architecture} \label{chap:Architecture}

The proposed architecture, as shown in Fig.~\ref{architecture}, consists of three main components: the IDS, the 5G Network Environment, and the Model Training Environment. The IDS utilizes advanced DPI integrated with ML models to enhance network security. DPI performs a detailed inspection of network packets, extracting essential data and traffic features, which can be leveraged by ML algorithms to accurately detect abnormal network behavior and malicious activity with high accuracy, offering more effective and adaptive protection for modern industrial communication systems. The 5G Network Environment, based on a real-world 5G Standalone Campus Network, enables seamless industrial application deployment while supporting more realistic attack simulations. It enables customizable attack scenarios and application behaviors, essential for training ML models with diverse datasets. By utilizing a real Private 5G Network instead of a simulated environment, the architecture ensures that the network traffic accurately reflects real world conditions, generating high quality and realistic datasets to enhance the accuracy and resilience of trained ML models. Meanwhile, the Model Training Environment includes advanced training servers for developing and optimizing detection models. Additionally, user-plane traffic is captured and forwarded from the Private 5G Network to the IDS via a dedicated mirror interface of the 5G core, or it can be securely transmitted unidirectionally to the Public 5G Network via a self-developed data diode \cite{songdatadiode, songdatadiode2} for further analysis or monitoring purposes. The architecture also includes a Database and Data Engineering node, which processes and structures the collected traffic, and a Visualization node that provides graphical insights into the detected results.


\vspace{2\baselineskip}

\section{Testbed Implementation} \label{chap:Testbed}

This section presents the testbed designed to validate the proposed architecture in detail.

\subsection{Application Cell using OPC UA Protocol}
The widely adopted industrial communication protocol OPC UA is selected for initial application deployment within the testbed. The Application Cell represents an automated warehouse, where packets are loaded onto and unloaded from shelves automatically. The Factory IO application is used to simulate hardware components at the field level, while the logic control program and OPC UA instances are deployed on industrial Programmable Logic Controllers (PLCs), called Revolution Pi (RevPi). The first RevPi module is connected to Factory IO through IO modules (IOs) to collect input data from and send output commands to the simulated field level components. An OPC UA server integrated into the RevPi module acts as a gateway, ensuring data accessibility for control programs and monitoring applications. The PLC program on the second RevPi is responsible to remotely control the warehouse application through an OPC UA client connected to the OPC UA server. Two 5G HAT modems equipped with SIM8200EA-M2 multi-band modules are employed to connect the RevPi devices to the Private 5G Network, ensuring high reliability and low latency communication for the application.

\subsection{Attack Cell}	
The Attack Cell generates attack traffic using TRex, an open source packet generator from Cisco, installed on a Dell Precision 3630 workstation equipped with an 10Gb X550-T2 network card. TRex provides high-performance traffic generation capabilities and is well-suited for simulating complex attack scenarios. Meanwhile, the 5G router Scanlance M800 connects the Attack Cell to the Private 5G Network, enabling efficient transmission of attack traffic within the testbed.

\subsection{Private 5G Network}
A Standalone Private 5G Network, installed by Telekom Deutschland GmbH on the campus of Helmut Schmidt University -- University of the Federal Armed Forces in Hamburg, serves as the backbone of the testbed. The network is based on the EDAV-I solution developed by Ericsson, which is compliant with 3GPP Release 16 and operates in the 3.7\,$-$\,3.8\,GHz frequency band \cite{HenryMeasurements}. The integrated mirror interface in EDAV-I captures and provides user-plane traffic with a bandwidth of up to 10 Gb/s, enabling robust and comprehensive data collection for further analysis.

\subsection{IDS}
The IDS is implemented on a cluster consisting of a R750xs and a R660xs Dell server, installed with the R\&S®PACE 2 library to enable real time protocol classification and packet analysis up to OSI Layer 7. Additionally, an external tool has been developed and integrated into the library to extract and enrich traffic features used for training and evaluation of detection models. These processes are performed on the campus High Performance Computing (HPC) cluster \cite{HSUper2024}, ensuring efficient handling of large datasets and accelerating processing. Furthermore, a Grafana dashboard is implemented to provide intuitive visualization of traffic patterns and detection results.

\section{Evaluation} \label{chap:Evaluation}

In this section, HEL flooding attacks that exploit the connection handshake mechanism of the OPC UA protocol are implemented as representative attack scenarios within the testbed. These attacks are carried out over untrusted, unencrypted packet streams and are based on findings from the Federal Office for Information Security (BSI) report \cite{BSI2017} and the study by Neu et al. \cite{Neu2019}. Statistical features calculated from network traffic generated within the testbed are also analyzed to assess their suitability for ML-based security solutions.

\subsection{Attack Simulation} \label{chap:AttackSimulation}	

The attack can be executed with adjustable attack parameters: HEL packets per flow \texttt{HEL\_pF}, flow multiplier \texttt{m}, and flow creation duration \texttt{d}, using the command \texttt{start -f HEL\_attack.py -m X -d Y}. The \texttt{HEL\_pF} parameter can be modified in the \texttt{HEL\_attack.py} script. Upon execution, the TRex server attempts to generate \texttt{X} attack flows per second for \texttt{Y} seconds, resulting in \texttt{X × Y} total attack flows. Using the \texttt{HEL\_attack.py} script, each flow follows three steps: (1) establishing a TCP connection to the OPC UA server with a unique source port, (2) sending \texttt{HEL\_pF} HEL packets, and (3) closing the connection after transmission. The implemented attack operates in a stateful traffic context, meaning its duration may vary depending on network quality and OPC UA server processing capacity.

\begin{figure}[b]
	\centering
	\includegraphics[width = 0.75\columnwidth]{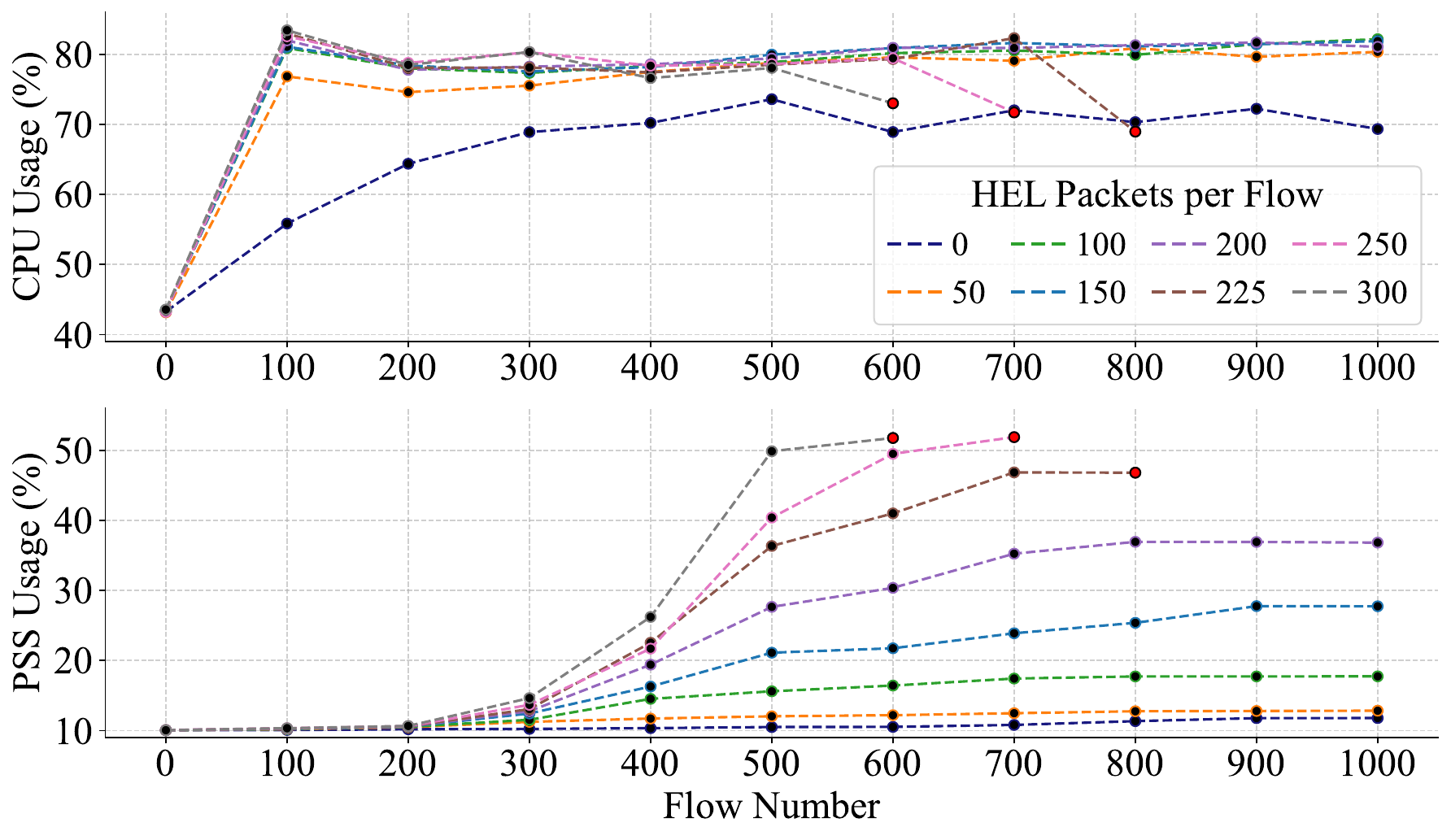}
	\caption{Impact of HEL flooding on OPC UA server resource utilization. Red points indicate termination of the server. Each data point represents the average of measurements from three independent attacks with identical attack parameters.} 
	\label{attack_results} 
\end{figure}

Fig.~\ref{attack_results} demonstrates the attack impact on OPC UA server resource utilization across varying attack parameters, with the flow creation duration \texttt{d} set to $1$\,s. CPU utilization metric represents the average CPU load, while PSS utilization metric indicates the peak Proportional Set Size (PSS) memory allocation, both measured either during the attack or about a $10$\,s interval, when no attack is performed (at Flow Number $ = 0$). During each execution, measurements are taken every $200$\,ms within the measured intervals. As shown in Fig.~\ref{attack_results}, CPU utilization increases significantly with the flow number and stabilizes around $80$\%, except when no HEL packets are sent, until the OPC UA server is terminated. Meanwhile, PSS utilization rises sharply as both the flow number and HEL packets per flow increase. These results validate the testbed capability to effectively simulate the attack scenarios within an industrial Private 5G Network. They also indicate that the combination of application- and transport-layer message flooding can significantly amplify the attack impact, emphasizing the need for DPI-based security solutions to deeply inspect network traffic and effectively detect application-layer threats.

\subsection{Traffic Evaluation} \label{chap:TrafficEvaluation}

\begin{figure}[t]
	\centering
	\includegraphics[width = 0.75\columnwidth]{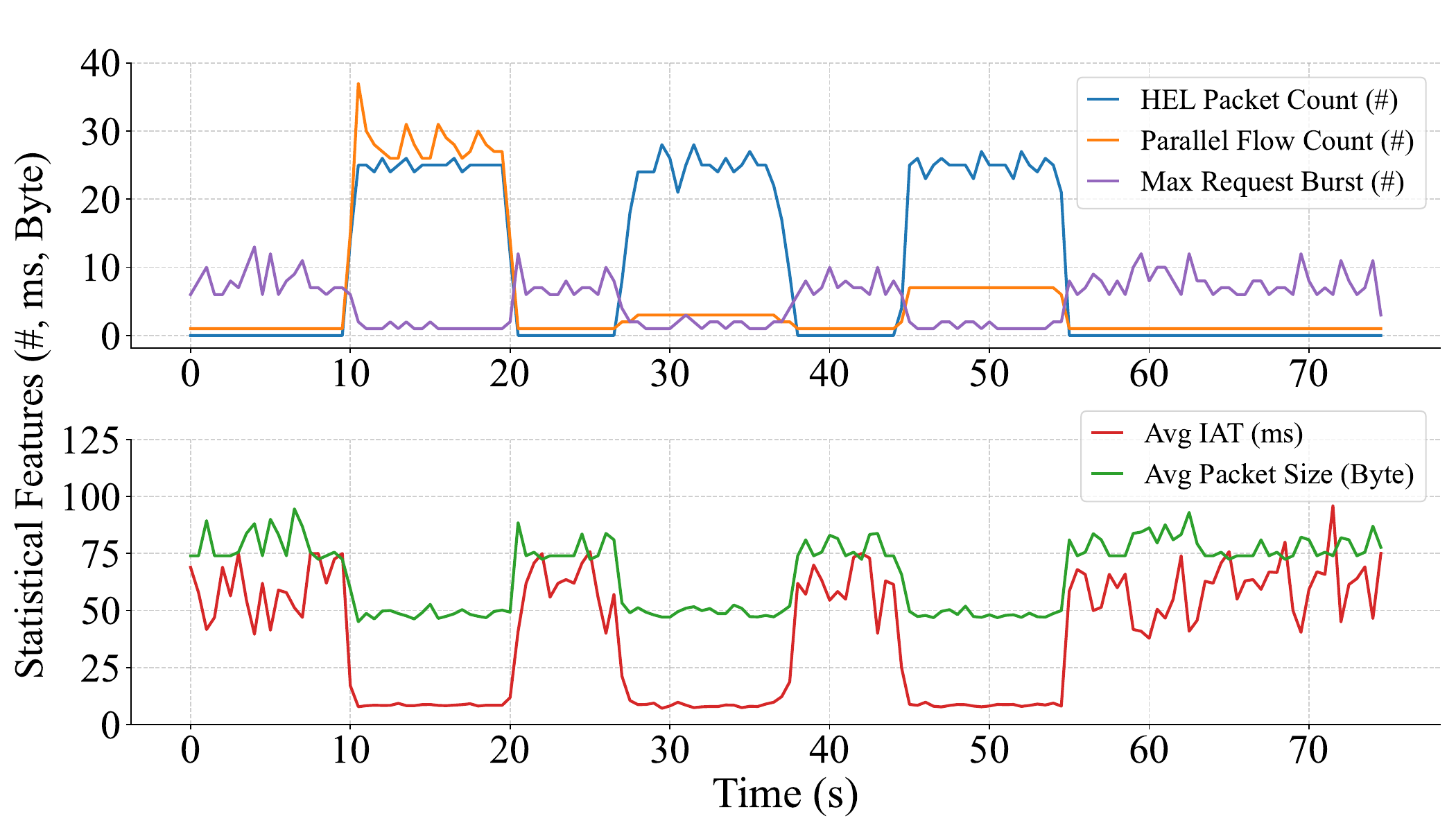}
	\caption{Statistical features were calculated in 500\,ms time windows during three example DoS attacks against the OPC UA server. The first attack ($\approx$\,9\,s\,–\,$\approx$\,21\,s) uses \texttt{HEL\_pF}\,$=$\,$1$, \texttt{m}\,$=$\,$50$; the second ($\approx$\,26\,s\,–\,$\approx$\,38\,s) uses \texttt{HEL\_pF}\,$=$\,$50$, \texttt{m}\,$=$\,$1$; and the third ($\approx$\,44\,s\,–\,$\approx$\,55\,s) uses \texttt{HEL\_pF}\,$=$\,$5$, \texttt{m}\,$=$\,$10$.}
	\label{evaluation} 
\end{figure}

To further analyze network traffic characteristics, 5G traffic of the testbed is forwarded to the IDS, where packets are thoroughly inspected and statistical features reflecting flow- and packet-level behavior were calculated. Fig.~\ref{evaluation} presents five selected features, based on insights from the study by Hindy et al. \cite{ML_Based_IoT}, Panigrahi et al. \cite{cicid2017} and observations of OPC UA traffic, namely HEL Packet Count, Parallel Flow Count, Max Request Burst, Average Inter-Arrival Time (IAT), and Average Packet Size. These features are computed from 5G traffic under both normal conditions and simulated DoS attacks using three representative attack parameter configurations. As illustrated, the extracted features reflect clear differences in traffic behavior between normal and attack scenarios, indicating their potential to support the development of ML-based security solutions. However, some features are more effective for specific attack strategies, while others are broadly applicable but sensitive to the context. For example, Parallel Flow Count is primarily effective in detecting DoS attacks that involve many concurrent flows, while Average IAT is sensitive to sudden bursts in traffic and is potentially less reliable under stealthy attacks or in the presence of network jitter. This underscores the importance of combining multiple features to increase the robustness and reliability of ML-based anomaly detection within an industrial Private 5G Network. By leveraging flow-based data, malicious and normal traffic are accurately labeled, facilitating the creation of a high-quality dataset for training and evaluating ML models in future work.

\section{Conclusions and future work} \label{chap:Conclusions}

This paper presents a testbed and software architecture for enhancing security in industrial Private 5G Networks. It enables the development and evaluation of advanced DPI- and ML-based security mechanisms, while also supporting the generation of realistic traffic datasets for training and validating ML models. The implemented testbed demonstrates the feasibility of simulating attack scenarios within an industrial Private 5G Network. The DPI module can deeply analyze network traffic, extract data from individual packets, and compute statistical features within predefined time windows. Other architectural components have also been tested, exhibiting expected operation. These results provide an initial validation of the proposed design under controlled conditions, confirming their readiness for further development. 

In future work, we will scale up the OPC UA Application Cell and implement a wider range of attacks, as outlined in the BSI report \cite{BSI2017} and the study by Neu et al. \cite{Neu2019}. The testing process will include untrusted, unencrypted and trusted, encrypted packet streams to evaluate the OPC UA application under different security configurations. Beyond OPC UA, the testbed will be extended to incorporate additional industrial communication protocols and a diverse range of attack scenarios in industrial environments. Consequently, DPI will be applied more extensively in the next phase to support this broader testing scope. Additionally, ML model training will leverage HPC to accelerate processing, enable large dataset training and improve model accuracy. Finally, the architecture will be validated in real-world environments for robustness and effectiveness in securing industrial Private 5G Networks.

\section*{Acknowledgment}

The authors would like to thank ipoque GmbH, H. Beuster, K. Tebbe, J. Jockram and F. Mueller for their valuable support. 

\section*{Funding}
This research is funded by dtec.bw – Digitalization and Technology Research Center of the Bundeswehr. dtec.bw is funded by the European Union – NextGenerationEU (project “Digital Sensor-2-Cloud Campus Platform” (DS2CCP), \href{https://dtecbw.de/home/forschung/hsu/projekt-ds2ccp}{https://dtecbw.de/home/forschung/hsu/projekt-ds2ccp}).

\bibliographystyle{unsrtnat}
\bibliography{references}

\end{document}